\documentclass{comnet}
\usepackage{graphicx}
\usepackage{dcolumn}
\usepackage{bm}
\usepackage{float}
\usepackage{color}
\usepackage{amssymb}
\usepackage{tikz}
\usetikzlibrary{shapes}
\usepackage{xcolor}
\usepackage{subfig}
\usepackage[percent]{overpic}

\begin{document}
\title{Cascading Failures in Complex Networks}

\shorttitle{Cascading Failures in Complex Networks} 

\author{{
\sc Lucas D. Valdez}$^*$,\\[2pt]
Department of Physics, Boston University, Boston, Massachusetts, USA\\
{\email{$^*$Corresponding author}}\\[4pt]
{\sc Louis Shekhtman}\\[2pt]
Network Science Institute, Northeastern University, Boston, Massachusetts, USA\\
{$\;$}\\[2pt]
{\sc Cristian E. La Rocca}\\[2pt]
Instituto de
  Investigaciones F\'isicas de Mar del Plata (IFIMAR)-Departamento de
  F\'isica, FCEyN, Universidad Nacional de Mar del Plata-CONICET, Mar
  del Plata, Argentina.\\ Department of Physics, Boston
  University, Boston, Massachusetts, USA\\
{$\;$}\\[2pt]
{\sc Xin Zhang}\\[2pt]
College of Transport and Communication, Shanghai Maritime University, Shanghai, China.\\ Department of Physics, Boston University, Boston,
  Massachusetts, USA\\
{$\;$}\\[2pt]
{\sc Sergey V. Buldyrev}\\[2pt]
Department of Physics,
  Yeshiva University, New York, New York,
  USA.\\ Department of Management, Economics and Industrial
  Engineering, Politecnico di Milano, Milano, Italy.\\
  Department of Physics,
  Boston University, Boston, Massachusetts, USA\\
{$\;$}\\[2pt]
{\sc Paul A. Trunfio}\\[2pt]
Department of Physics, Boston University, Boston,
  Massachusetts, USA\\
{$\;$}\\[2pt]
{\sc Lidia A. Braunstein}\\[2pt]
Instituto de
  Investigaciones F\'isicas de Mar del Plata (IFIMAR)-Departamento de
  F\'isica, FCEyN, Universidad Nacional de Mar del Plata-CONICET, Mar
  del Plata, Argentina.\\
  Department of Physics, Boston University, Boston,
  Massachusetts, USA
{$\;$}\\[2pt]
{\sc and}\\[4pt]
{\sc Shlomo Havlin}\\[2pt]
Department of Physics, Bar Ilan University, Ramat Gan, Israel.\\
Department of Physics, Boston University, Boston,
Massachusetts, USA.\\
Tokyo Institute of Technology, Yokohama, Japan
{$\;$}}

\maketitle

\begin{abstract}
{Cascading failure is a potentially devastating process that spreads on
real-world complex networks and can impact the integrity of
wide-ranging infrastructures, natural systems, and societal
cohesiveness. One of the essential features that create complex network
vulnerability to failure propagation is the dependency among their
components, exposing entire systems to significant risks from
destabilizing hazards such as human attacks, natural disasters or
internal breakdowns.  Developing realistic models for cascading failures
as well as strategies to halt and mitigate the failure propagation can point to new
approaches to restoring and strengthening real-world networks. In this review, we
summarize recent progress on models developed based on physics and
complex network science to understand the mechanisms, dynamics and
overall impact of cascading failures. We present models for cascading
failures in single networks and interdependent networks and explain how
different dynamic propagation mechanisms can lead to an abrupt collapse and
a rich dynamic behavior. Finally, we close the review with novel
emerging strategies for containing cascades of failures and discuss open
questions that remain to be addressed.}
{Cascading Failures, Complex Networks, Spatial Networks, Network of Networks, Percolation, Network Robustness}
\end{abstract}
  
\section{Introduction}

It is well accepted that many complex systems can be represented,
analyzed and better understood as networks. Examples include World
Wide Web, social, and related online networks; the power grid,
Internet, traffic, airline and related infrastructure networks; and
neural and physiological networks~\cite{newman2018networks}. Formally,
a network or graph is a set of nodes or vertices connected via
internal links. Despite the network's mathematical simplicity, it also
establishes a powerful and versatile tool to characterize and
understand many complex systems in nature, technology, and society. In
a node-link network structure, one of the main measures characterizing
topology is the degree distribution $P(k)$ which gives the probability
that a node has $k$ connections (links). Networks with a
characteristic average degree or connectivity can be regarded as
homogeneous, whereas networks that lack a characteristic degree and
instead consist of a broad range of degrees can be regarded as
heterogeneous networks. A typical example of a homogeneous network is
the Erd\H{o}s R\'enyi network which has a Poisson degree distribution,
while for heterogeneous networks a typical example is the Scale-Free
(SF) network, with a power law degree distribution with exponent
$\lambda$. Another important feature of real-world networks is that
the nodes are not randomly connected but display nonrandom
patterns of connections with features such as clustering, degree-degree correlations
and modularity~\cite{newman2018networks,barrat2008dynamical}.

In our everyday world, a large number of processes occur on top of
networks, such as the spread of diseases in contact networks, opinion
formation in social networks and synchronization of neurons in the
brain~\cite{newman2018networks,barrat2008dynamical}.  One of the most
dramatic processes that spread on complex networks is the cascade of
failures when a failure in part of the system leads to further
failures in the same and other systems which then continue to
propagate. Eventually, the entire system could become dysfunctional
and catastrophically collapse. For example, in an interdependent
system such as the organs of the human body, the malfunction of the
nervous system or another physiologic system can affect various organs
generating negative feedback that could lead to system collapse and
death. Similarly, in an electrical power grid, the damage of a node
could trigger an overload on other components of the grid that further
propagates to the entire system that could lead to a catastrophic
collapse of the
system~\cite{dobson1992voltage,carreras2016north,motter2002cascade}. The
initial failure that triggers the cascade could be due to human
attacks or natural disasters, such as the tsunami in Japan in 2011,
and the forest fire in California in August 2018. Hurricane Katrina in
2005 is another example of extreme event with catastrophic effects on
several
systems~\cite{chang2014infrastructure,leavitt2006infrastructure,comfort2006communication}.
In addition, studying cascading failures can explain why networks evolve towards topologies that attenuate these cascades. For example, for the Escherichia coli,
Ref.~\cite{klosik2017interdependent} showed that for a gene regulatory
domain (network) which depends on a metabolic domain through a protein
domain, small perturbations originated in the metabolic domain trigger
smaller cascades than those originated in the gene regulatory
network. The higher robustness of the metabolic network is explained
because it is more coupled with the environment than the other
domains, and hence, the metabolic network needs to be more robust
under external fluctuations. On the other hand,
Ref.~\cite{reis2014avoiding} showed for networks obtained from
functional magnetic resonance imaging (fMRI) experiments and
simulations, that different modules in the functional brain network
have correlations in their node connectivities which increase the
robustness of the system against cascading failure.

Studying cascades of failures in networks is not only critical for
understanding how the network structure impacts its resilience to
catastrophic cascades, but more importantly it helps to develop tools
that can predict, mitigate, prevent and recover the system from such
failures. Physics and network science have an essential role in
understanding cascading failures since tools from statistical
mechanics, such as phase transitions, percolation theory and nonlinear
dynamics, are useful to describe and comprehend the process, and could
help in developing strategies to halt or mitigate the catastrophic
effects of collapse. Here, we review the main physics models used to
improve our understanding of the dynamics of cascading failures. We
must emphasize that these physics models are simplified models of real
infrastructures and their predictions serve to demonstrate mechanistic
possible behaviors of systems, rather than provide the type of precise
predictions for a particular system as would be done in
engineering. Any studies of real interdependent infrastructures must
take into account technical details of the system under investigation
and thus must depend on a large number of parameters. This makes it
very difficult to gain any mechanistic insight into the actual
underlying processes leading to the observed behaviors, which is the
goal of physicists.

\section{Cascade of Failures in Single Complex Networks}

The study of dynamic spreading models in single or isolated networks
allows us to discover how different types of internal dynamics affect
the system functionality or overall robustness. Dynamic processes can
occur in isolated networks and propagate via cascades, i.e., processes
in which a component that becomes dysfunctional lead to other
components that depend on it (directly or indirectly) to also become
dysfunctional. The origin of this cascade has several possible sources
and can, for example, be due to: (i) the redistribution of the load on
a node or link leading to overloads as the model of
Ref.~\cite{motter2002cascade}, (ii) the existence of direct
dependencies in which if a node becomes dysfunctional then all the
nodes that depend on it also become
dysfunctional~\cite{parshani2011critical}, or (iii) the number of
functional neighbors is above or below a threshold as in bootstrap and
k-core percolation,
respectively~\cite{dorogovtsev2006k,baxter2010bootstrap,baxter2011heterogeneous,di2018insights}. Note,
that the overload models are in many ways similar to the models with
direct topological dependencies: node or link A, whose failure was
caused by the overload due to redistribution of load after failure of
node or link B, dynamically depends on B, because A cannot function
after B fails. However, the major distinction between the overload and
topological models is the dependence of the former on the sequential
order of failures, while for the latter the final outcome depends only
on the topology of network. The vulnerability of single networks to
the cascade of failures can be seen in an abrupt transition at a
critical fraction of functional nodes.  While the motivation of some
of these models is to study failure propagation in infrastructure
systems, they could be applied to other processes that propagate as
cascades, such as, activation process in living neural networks in
which cascades emerges because neurons fires under external
stimulation or if they receive fires from their input
neurons~\cite{breskin2006percolation,soriano2008development,monceau2018neuronal}. Another
possible application is the study of information propagation in social
networks, like the propagation of a hashtag or meme in Twitter, since
real data suggested that individuals transmits the information as a
complex contagion process where an individual needs multiple inputs
from several neighbors to transmit the
information~\cite{monsted2017evidence,romero2011differences}.  In this
section, although not exhaustive, we present models of cascades of
failures in single networks and some of the most recent insights.

\begin {figure}[H]
\vspace{0.3cm}
\centering
\includegraphics[scale=0.5]{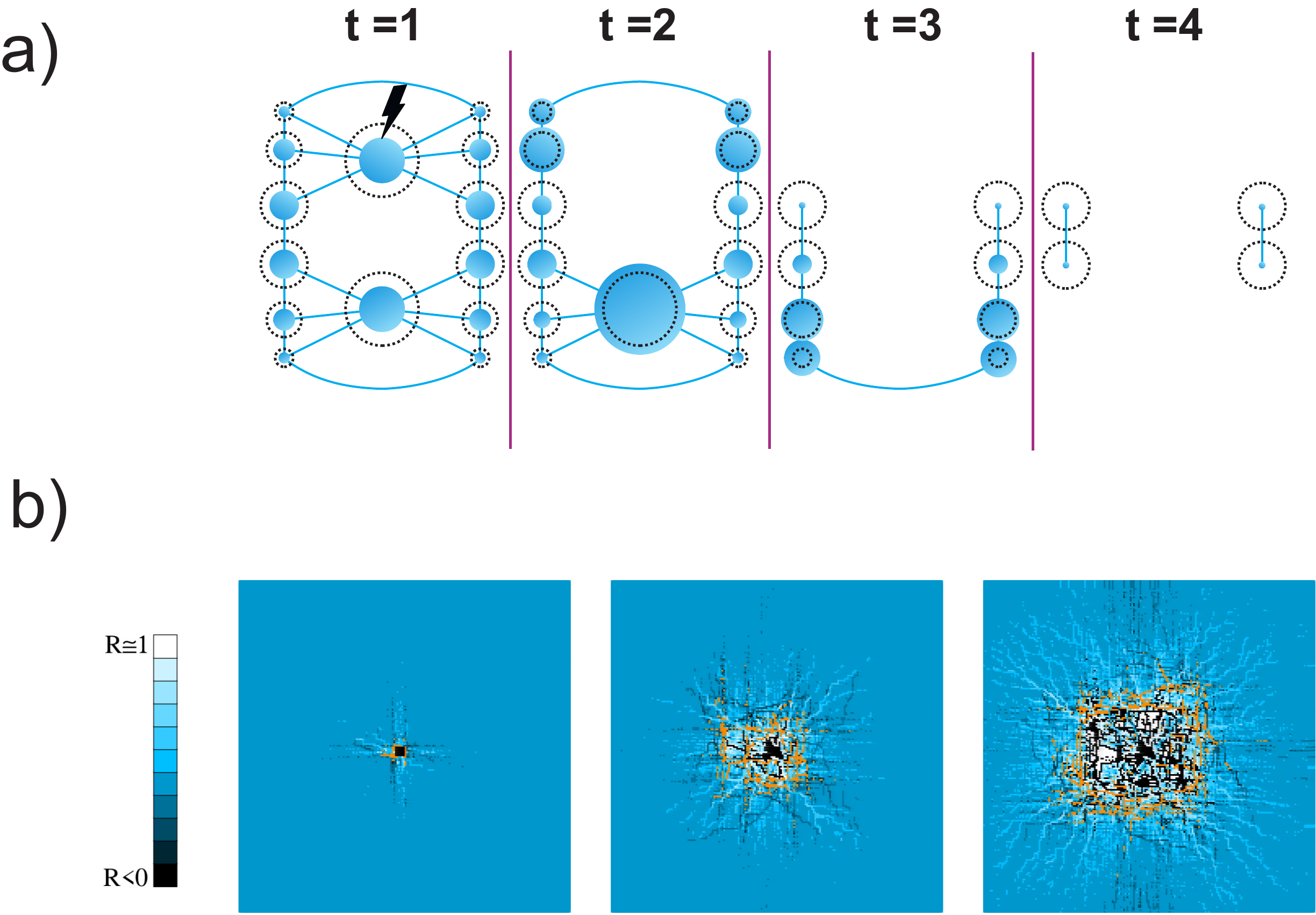}
\vspace{0.3cm}
\caption{Panel a): schematic figure of the time evolution of the
  overload model presented in Ref.~\cite{motter2002cascade}. The length of the
  radius of the node (blue circle) represents its load, i.e., the
  number of shortest paths that pass through that node, while the
  radius of the black dashed circle represents its capacity. In step $t=1$,
  one node fails which will change the load of the rest of nodes. In
  the following steps, nodes fail if their load exceed their capacity
  and they are removed. Panel b): Time evolution of the propagation of
  the overload failures in a lattice under localized attack at the
  center~\cite{zhao2016spatio} for time $t=2$, $t=4$, and $t=6$ (from
  left to right). The parameter $R$ represents how much free capacity
  a node has available ($R\approx 1$ corresponds to a node without
  load, and $R<0$ corresponds to a failed node because its load
  exceeded its capacity). For each snapshot at time $t$, the nodes
  that fail at that time are in orange.}\label{fig.overl}
\end{figure}

A simple overload model was proposed in Ref.~\cite{motter2002cascade}
to study the cascade of failures triggered by the initial failure of
one node. In this model, each node has a fixed maximum capacity to
support load which is given by node betweenness, i.e. the number of
shortest paths that pass through that node. This maximum capacity is
defined as the initial load of the node multiplied by $(1+\alpha)$,
where $\alpha>0$ is called tolerance. If a node's load exceeds its
maximal capacity, then the node fails and the shortest paths change,
generating subsequent overloads and failures and until the system
reaches a steady state (see Fig.~\ref{fig.overl}a). For heterogeneous
networks, if the initial failure is random, the system is so robust
that cascades of failure usually do not arise. However, if the node
with the highest connectivity or the highest load is the first to
fail, then a cascade is triggered, and the size of the functional
network significantly shrinks. This model was extended on embedded
two-dimensional Euclidean networks to understand how a localized
attack (in which a randomly chosen node and several shells around it
are removed) leads to propagation of cascading failures through the
network~\cite{zhao2016spatio} (see Fig.~\ref{fig.overl}b). In these
networks, a universal behavior of the failure propagation has been
observed, since in different network substrates, the overload
propagates at a nearly constant velocity which increases as the
tolerance of the system decreases. In this research, the authors
suggest that cascading overloads can be mapped into dependency
cascading failures. A similar propagation has also recently been
observed for the models of traffic jams~\cite{zhang2019scale}.

Recently, the model of Ref.~\cite{motter2002cascade} has been explored
for a case when a macroscopic fraction $1-p$ of nodes fails at the
initial condition~\cite{kornbluth2018network}, as opposed to one node
in the original model~\cite{motter2002cascade}.  In this case, the
order parameter can be taken either as the fraction of survived
nodes, or as the fraction of nodes in the giant component (GC). At the
end of the cascade, the system undergoes a first-order transition on the
$\alpha,p$ plain for small $\alpha$. However, as the tolerance
increases, and $p$ decreases the transition becomes continuous as in ordinary
percolation.

Similar threshold
models~\cite{dobson1992voltage,carreras2002critical,carreras2004complex}
have been used to simulate cascading failures in a more realistic
direct current approximation of the electric power grid, in which the
nodes are connected with transmission lines of given resistance and
the nodes can be of three types: generators adding current to the
system, loads taking current away from the system, and transmission
nodes, which conserve current. Each node satisfies the Kirchhoff
equation. At the beginning, a line or a fraction of lines is
destroyed. As a result, the current redistributes, and if the current
in a line exceeds a certain threshold, established by the N-1
criterion~\cite{ren2008long}, the line also fails. This process leads
to cascading failures, in which the distribution of the blackout sizes
follows an approximate power law as supported by the empirical
statistics~\cite{carreras2016north}.  This and earlier observations and
modeling of the cascades in power grids lead in
Refs.~\cite{carreras2000initial,carreras2001evidence,carreras2004evidence,newman2011exploring,dobson2007complex,carreras2004complex,ren2008long,dobson2012estimating,kim2012estimating,qi2013towards}
to a conjecture that the power grids drive themselves to the critical
point like the self-organized criticality models~\cite{bak1993punctuated,maslov1995exactly,
  maslov1995time,paczuski1996avalanche}.

However, similar models~\cite{spiewak2016study, pahwa2014abruptness}
display the same phenomenology as in the model of Ref.~\cite{motter2002cascade} based on
betweenness, showing a bimodal distribution of the blackout sizes. The
reason for this difference is that these models follow different rules
of cascade propagation. In the former models, at each stage of the
cascade, only one of the overloaded lines is removed after which the
currents in rest of the lines are recalculated, while in the latter
all overloaded nodes are removed at once. This difference illustrates
a crucial difference between the overload models in which the outcome
depends on the dynamics of the cascade and the topological models, in
which the outcome is independent of the dynamics.  The seminal paper
of Hines et al.~\cite{hines2010topological} who include in their model
dynamical effects also observed the difference in behavior of cascades
in topological models of failure and the models based on overloads.

The importance of transient dynamics in the cascade of failures was
also demonstrated in
Refs.~\cite{schafer2018dynamically,yang2017cascading}.  In
Ref.~\cite{schafer2018dynamically}, it was observed that after the
failure of a single link, the nodes at a distance closest to the
initial failure tend to fail first, where distance is defined as the
smallest weighted shortest path. This finding sheds new light on the
understanding of propagation patterns on networks by measuring the
speed of a cascade.

Another direction of research on cascade of failures in power grids
involves identifying those variables that are good predictors for the
final size of a cascade. It was found that one of the predictive
factors is the vulnerability of a component, i.e., the probability that it
fails, and another predictive factor is the co-susceptibility which is
the tendency for a group of components to fail
together~\cite{yang2017vulnerability}. This phenomenon is similar to the
concept of dependency groups that were modeled in single
networks~\cite{parshani2011critical} and interdependent
networks~\cite{wang2018group}. Another concept that has been proposed to
study for the cascade of failures in power grids is the ``vulnerable set''
which is defined as the set of lines with vulnerability above a given
threshold value~\cite{yang2017small}. Using this measure and through
stochastic simulations of the United States and South Canada power grid
network, it was found that a small fraction of the transmission lines in
the network is vulnerable to malfunction or physical damage, called
``primary failure''. Moreover, the topological and geographical
distances of the initial failure to this vulnerable set determines the
amount of cascading failures, as measured by the reduction in the amount
of power delivered to consumers.

While many models describing the cascade of failures in single networks
correspond to irreversible processes, in practice, a node can often
return to its previous state. These systems could describe, for example,
neural networks in which there are inhibitory and excitatory neurons
leading to cascades of activation or deactivation of
neurons. Introducing the concept of recovery in a model,
Ref.~\cite{majdandzic2014spontaneous} studied nodes that can fail either
spontaneously or if their number of active neighbors are below a
threshold, and included the ability of nodes to recover, i.e.,
return to the active state after a period of time. They found that
the fraction of active nodes exhibit hysteresis and if the system size
is finite then the system flips between a state with a majority of
active nodes and a state with an inactive majority. In
Refs.~\cite{valdez2016failure,bottcher2017failure} it was found that
this model also exhibits oscillations in the fraction of active and
inactive nodes for some values of the control parameters. Similarly, for
a model of neural networks in complex networks, the system can exhibit
hysteresis in the fraction of active neurons, as well as oscillations
even for a network with a small number of
nodes~\cite{goltsev2010stochastic}.

In the following section, we review several studies of cascade of
failures in interdependent networks showing that dependency links
between networks lead to
cascading failures between networks, rich dynamics and even to abrupt
transitions.

\section{Cascade of Failures in Interdependent Networks}\label{sec.InterComp}

Before 2010, researchers primarily studied processes on single
networks. However, real-world networks such as the infrastructure
systems are not isolated but depend on one
another~\cite{rinaldi2001identifying,vespignani2010complex}. Crucially,
if a small failure occurs in one system it could lead to other
failures and propagate, leading to the complete collapse of the system
of systems. Such systems of systems can be represented as
interdependent networks composed of individual networks that depend on
each other for functionality. Within each individual network, the
links between nodes are {\it connectivity} links, while between the
networks {\it dependency} links connect the nodes of different
networks. In interdependent networks, the cascading failure consists
of a feedback mechanism in which the failure propagates among the
networks back and forth through dependency links until the system
reaches the steady state.
\begin{figure}[h]
  \centering
  \includegraphics[width=200pt]{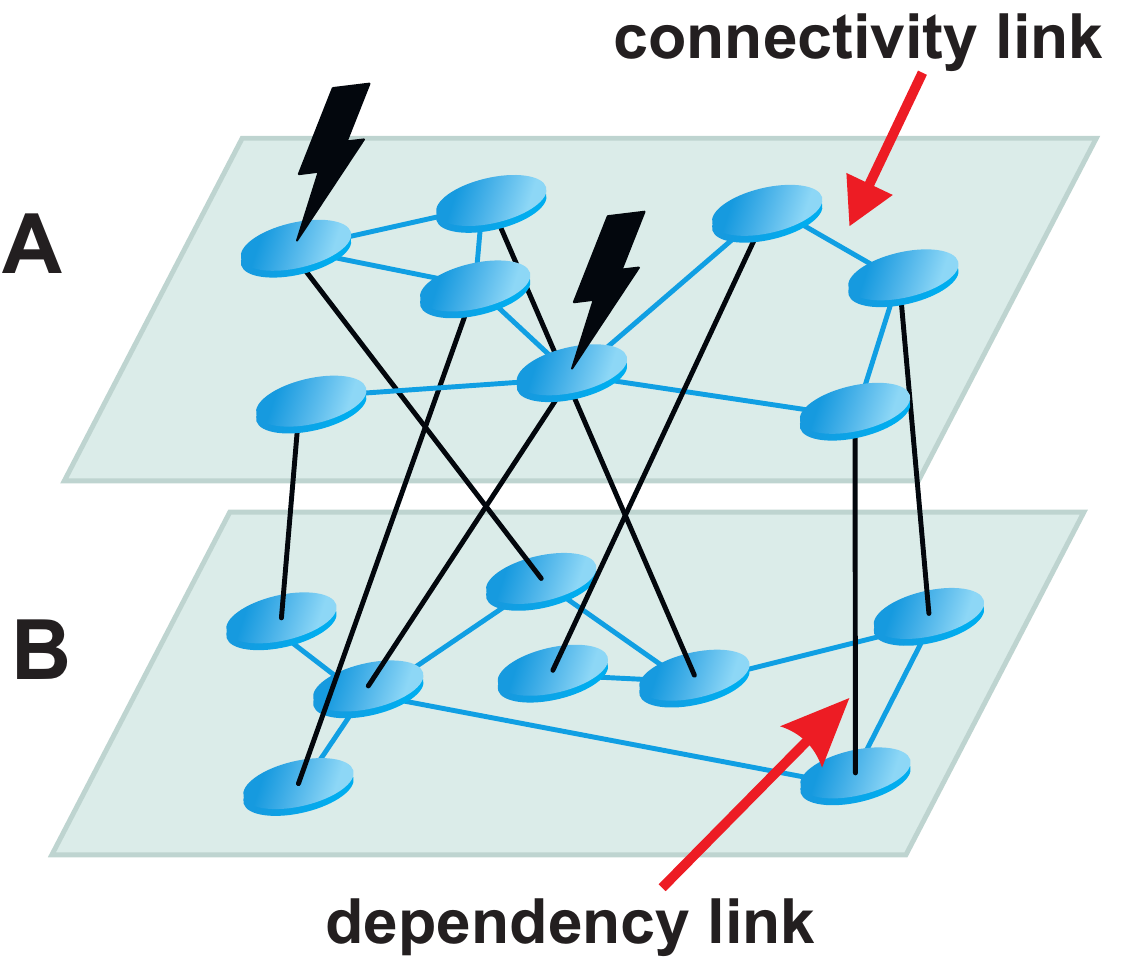}
  \caption{ Schematic
    representation of two interdependent networks A and B with
    one-to-one interdependency where two nodes in network A fail
    externally (represented by lightning). Blue links are {\it connectivity}
    links and black links are {\it dependency} links.}\label{fig.Princ}
\end{figure}

In these interdependent networks, the functionality of nodes in one
network depends on the state of the nodes in other networks. In these systems, a failure in one
network may trigger a domino effect which can result in the collapse
of the whole system.  For example, the great blackout of Italy in 2003
demonstrated that breakdowns in power grids strongly impact other
systems such as communication and transportation networks, and the
failure of these networks, in turn, accelerates the failure of the
power grid~\cite{rosato2008modelling}. The modeling of the propagation
of this cascade of failures across several networks has received broad
interest in recent
years~\cite{buldyrev2010catastrophic,d2014networks,bianconi2018multilayer}
because many real-world systems are not isolated but depend on
others. Such interdependency can make the system more vulnerable
compared to single isolated networks. The manifestation of this
vulnerability is seen in the propagation of failures as a cascade and
the emergence of an abrupt collapse: in isolated networks, the
percolation transition is a continuous second order transition while
in an interdependent system it is a hybrid phase
transition~\cite{baxter2012avalanche,zhou2014simultaneous}. For
interdependent systems near the transition point, the system is
metastable: very small additional damage can cause the complete
collapse of the system, which otherwise appears stable. 

\begin {figure}[H]
\vspace{0.3cm}
\centering
\includegraphics[scale=0.5]{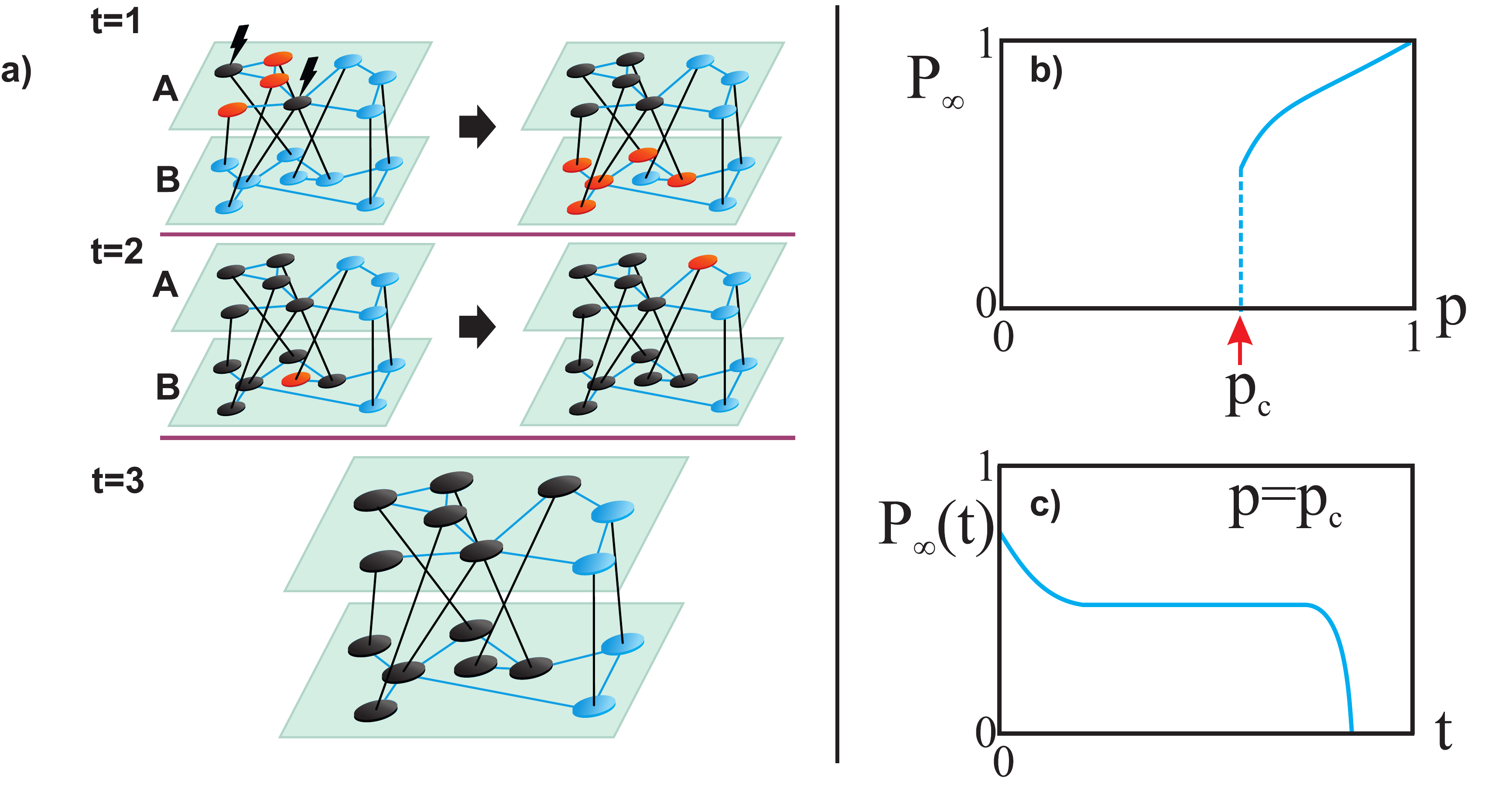}
\vspace{0.3cm}
\caption{Panel (a): Schematic illustration of the temporal evolution
  of the cascade of failures for the model presented in
  Ref.~\cite{buldyrev2010catastrophic}. The black nodes are
  dysfunctional, blue nodes are functional and red nodes will become
  dysfunctional because they belong to finite clusters or fail due to
  the interdependence. In step $t = 1$, on the left, two nodes fail,
  so network $A$ separates into three clusters. The two smallest
  clusters will become dysfunctional which will cause the nodes of
  network B to fail in the next time step (see the interdependent
  network on the right for $t=1$). At $t = 2$ the nodes of network B
  fail. At $t = 3$ the steady state is reached. Panel (b): Schematic
  figure of the fraction of nodes in the GC, $P_{\infty}$, when a
  fraction $1-p$ is initially removed. The red arrow indicates the
  value $p_c$ at which the system undergoes cascading failures and abrupt collapse. Although
  the transition is abrupt, it is different from a normal abrupt
  transition since just above but close to the threshold point, the
  fluctuations of the sizes of the GC diverges as a power
  law~\cite{baxter2012avalanche, serrano2015escaping}. Panel (c):
  Schematic figure of the time evolution of the fraction of nodes in
  the GC, $P_{\infty}(t)$, at $p=p_c$. Before the system collapses,
  $P_{\infty}(t)$ has a long plateau which increases when the system
  size increases~\cite{buldyrev2010catastrophic,zhou2014simultaneous}.}\label{fig.steps}
\end{figure}

A simple model of interdependency between networks was developed
by Buldyrev {\it et al.} in 2010~\cite{buldyrev2010catastrophic} and
showed that such systems could undergo abrupt collapse under random
failures.  In this study, the system consists of two interdependent
random networks $A$ and $B$, in which nodes in one network depend via
a one-to-one correspondence on nodes in the other network and vice
versa (see Fig.~\ref{fig.Princ}). The functionality of the network
depends on internal and external rules~\cite{di2017cascading} of failure propagation. The
internal rules govern the conditions that a node in a network will
fail exclusively due to the states of the nodes in the same
network. On the other hand, the external rules indicate under which
conditions a node fails due to the states of the nodes in the other
networks. The internal rule of failure for the particular process described in Ref.~\cite{buldyrev2010catastrophic} (called mutual percolation) is that nodes
that do not belong to the GC of each network fail, while the external
rule of failure is that if a node fails in one network, its
interdependent node in the other network also fails. The process of
cascade of failures begins with a random failure of a fraction of
nodes in network $A$. This failure generates finite clusters that do
not belong to the GC which fail and whose failures propagate to
network $B$ through the dependency links. Subsequently, finite
clusters in network $B$ become dysfunctional and propagate the failure
back to network $A$ and so on. This cascading failures process is
repeated until the system reaches a steady state in which there are no
more finite clusters (see illustration in Fig.~\ref{fig.steps} a) and
the sizes of both GC are the same. In the thermodynamic limit,
Buldyrev {\it et al.}~\cite{buldyrev2010catastrophic} showed that in a
random network there exists a threshold value of the initial failure
$1-p_c$, at which the GC in both networks collapses abruptly since
they cannot sustain each other (see Fig.~\ref{fig.steps}b).  Around
the transition point, the time evolution of the size of the GC undergoes
a plateau stage in which it decreases very slowly, and after that, the
dilution accelerates and the system collapses (see
Fig.~\ref{fig.steps}c). The importance of this stage is that a useful
strategy could take advantage of this slow dynamic to stop the cascade
of failure using only a small number of resources.

One of the main advantages of the mutual percolation model is its
simplicity which, allows it to be studied using percolation
theory~\cite{buldyrev2010catastrophic,son2012percolation,stauffer2014introduction}. In
addition, Ref.~\cite{baxter2012avalanche} showed that interdependent
networks can be studied analytically in terms of avalanches. However,
we must emphasize that this model cannot be directly applied to the
interdependence of the power grid and Supervisory Control and Data
Acquisition (SCADA) system~\cite{rezai2017key} which was mentioned in
Ref.~\cite{buldyrev2010catastrophic} as an illustration only, because
the internal rule of failures in the power grid is not related to the
global connectivity. In fact, islanding, i.e. separating the power
grid into disconnected independent parts can be an efficient tool for
stopping cascade
propagation~\cite{spiewak2016study,korkali2017reducing}. Moreover,
Ref.~\cite{buldyrev2010catastrophic} ignores technical details of the
SCADA system~\cite{rezai2017key}.  Following
Ref.~\cite{buldyrev2010catastrophic}, several studies have
incorporated more realistic features into the original model and
investigated their effects on the cascade of failures, such as
degree-degree
correlations~\cite{zhou2012assortativity,valdez2013triple,buldyrev2011interdependent,min2014network,watanabe2014cavity,serrano2015escaping},
link
overlap~\cite{hu2013percolation,baxter2016correlated,cellai2013percolation,cellai2016message,min2015link},
clustering~\cite{huang2013robustness,shao2014robustness,tian2014robustness},
directed networks~\cite{liu2016breakdown,liu2019multiple}, multiple
interdependent
networks~\cite{gao2011robustness,gao2012networks,gao2013percolation,bianconi2015mutually,bianconi2014multiple},
multiple dependencies~\cite{di2017cascading,shao2011cascade}, spatial
embedding effects (will be discussed in the next
section)~\cite{li2012cascading,
  bashan2013extreme,danziger2014percolation,danziger2015interdependent},
hyperbolic interdependent networks~\cite{kleineberg2017geometric},
targeted and localized
attack~\cite{huang2011robustness,yuan2015breadth,baxter2018targeted}
and generalization of interdependence to dynamical
systems~\cite{danziger2018dynamic,scala2016cascades,zhang2018modeling}. Below,
we summarize some of the lines of research that have been developed in
the last years. Note that most of these extensions include the GC
internal rule and the one-to-one dependency.

{\it Multiple Interdependent Networks:}
Refs.~\cite{gao2011robustness,gao2012networks,gao2013percolation}
generalize the mutual percolation model to the system of more than two
independent networks, or Network of Networks (NON), focusing on how
different topologies at the global scale (of interdependence between
networks) affect the size of the GC and the percolation threshold. At
a global scale, networks are represented by ``supernodes,'' and a
``superlink'' between two supernodes represents all the dependency
links between these networks. In these primitive models of NONs the
dependency links establish a one-to-one correspondence (isomorphism)
between the nodes of all the networks in the NON, which means that if
a superlink connects two networks $A$ and $B$, this implies that every
node in $A$ has a dependency link towards a unique node in $B$ and
vice versa. In subsequent
works~\cite{baxter2012avalanche,son2012percolation,bianconi2015mutually},
it was explained that the mutual GC in these networks does not depend
on the global topology of the NON by arguing that networks with
isomorphism are equivalent to a single network with multiple layers of
links (called a multiplex), regardless of the global structure
(tree-like or with loops).

{\it Autonomizing Interdependent Networks:} In contrast to the
original model of interdependent networks
(Ref.~\cite{buldyrev2010catastrophic}), it is expected that
in real systems there exist nodes that are independent or autonomous
since, e.g., a communication system in SCADA may have battery
supplies, and hence it does not depend on any node of the power
grid. Ref.~\cite{parshani2010interdependent} studied the effect of
random autonomization in which a fraction $1-q$ of randomly chosen
nodes do not depend on the nodes of the other network. In the limit
$q=1$, the networks are fully interdependent as in
Ref.~\cite{buldyrev2010catastrophic} and the system experiences an
abrupt collapse, while for $q=0$ both networks are isolated and the
transition is continuous. They found that the transition could change
from discontinuous to continuous as the interdependency fraction is
reduced below a critical fraction $q_c$, which is analogous to the
critical point of the van der Waals phase diagram. Later work studied
the effect of targeted autonomization in which a fraction of nodes
with the highest degrees in each network are set as
autonomous~\cite{schneider2013towards,valdez2014triple}. They obtained
that the robustness of the GC is significantly improved compared to
random autonomization because nodes with the highest degrees tend to
maintain the GC connected. Moreover, Ref.~\cite{valdez2014triple}
showed that for heterogeneous networks, such as scale-free networks,
the system might have two transitions: one continuous and the other
discontinuous, and three different characteristic GC sizes with
thresholds that converge into a triple point, similar to the triple
point in the phase diagram of water. This third phase emerges because
as the initial failure of $1-p$ fraction of nodes increases, after the
first transition the system does not completely collapse, but is
composed of a core of hubs that were autonomized and mantains
the network's functionality.

{\it Modular Networks and Interdependent Hierarchical Networks:} While
many previous works on interdependent networks have been developed on
networks without internal structure, real systems have a community
structure at a mesoscopic scale such as the brain, social and
financial networks, and infrastructure across
cities. Ref.~\cite{shekhtman2015resilience} introduced one of the
first models of interdependent networks with communities in which
bridge nodes (see Glossary) are attacked first. For the case of two
interdependent networks and regardless of the number of communities,
they observed that the system undergoes at most two-phase
transitions. The first transition is due to the disconnection between
modules or communities (see Glossary) and the system splits into
several interdependent communities; whereas the second transition
corresponds to the collapse of the communities caused by the
dependency links. However, Ref.~\cite{shekhtman2018percolation} showed
that if communities are organized in a hierarchical structure with $n$
levels, such as the infrastructure at the levels of a city, state, and
country, the system decays abruptly, at most, at $n$ different values of
the initial damage. Each collapse corresponds to the disconnection of
a hierarchy level. This result indicates that at a mesoscopic scale,
the effect of community structure on the robustness of the system
saturates as the number of communities increases, but higher
organization levels of the network structure make the system more
fragile.

{\it Internal Rules of Functionality:} Several alternative internal
rules to the GC were explored such as k-core
percolation~\cite{di2017cascading,panduranga2017generalized,azimi2014k},
functional finite clusters~\cite{di2016cascading,
  yuan2017eradicating}, overload
model~\cite{zhang2018modeling}, and redundant
interdependencies~\cite{radicchi2017redundant}. For instance, in
Ref.~\cite{di2016cascading} the cascade of failures was studied for
interdependent networks in which a node does not fail internally if it
belongs to a cluster of size $s\geq s^*$, where $s^*$ is a
threshold. They found that depending on the value of $s^*$, the
transition is continuous or discontinuous. For small values of $s^*$
the transition tends to be continuous because in the limit of $s^*=1$
the system behaves as a single network. This shows that not only
interdependent links are necessary to generate an abrupt collapse, but
the rule of internal failure is crucial for a cascade of failure and
abrupt collapse.


{\it Stability:} On a global scale, the mean size of the GC at the
steady state is usually used as a measure of the stability of an
interdependent network. However, Ref.~\cite{kitsak2018stability}
pointed out that, conversely, from a microscopic point of view, the probability that a node belongs to the GC at the end of the cascade, may
be a good measure for the reliability of a node. For two
interdependent networks, the stability decreases significantly if
either of the two networks is homogeneous, and the most stable system
corresponds to two heterogeneous networks.  This phenomenon is due to
the hubs which have a higher probability of belonging to the GC and
serve as its anchors in a network~\cite{kitsak2018stability}, which
reaffirms the role of the hubs of preserving the GC as it was shown in
previous works~\cite{schneider2013towards,valdez2014triple}.

{\it Finite systems:} While many works have been developed in the
thermodynamic limit, real networks have a finite size of the order of
several thousand of nodes~\cite{coghi2018controlling} and hence, the
results obtained from these networks could be heavily affected by
finite size effects.  To solve this problem,
Ref.~\cite{radicchi2015percolation} developed a new approach based on
a non-backtracking matrix to study the robustness of real
interdependent networks that do not necessarily have a treelike
structure which is useful to compute the transition point. Another
work pointed out the importance of taking into account the
fluctuations in the size of the GC to characterize the robustness of
finite interdependent networks~\cite{coghi2018controlling}. For these
systems, they observed that the mode ~\cite{manikandan2011measures} of
the size of the giant component (given by the value of the size of the
GC with the highest probability among realizations) shows an abrupt
behavior, indicating that in finite networks this magnitude is a
better estimator of the abrupt nature of cascades of failures in real
systems.

{\it Generalization of Interdependence to Dynamic Systems:} One of the
main features of the original model of interdependent
networks~\cite{buldyrev2010catastrophic} that allows an exact theory
with generating functions in the thermodynamic limit is that the state
of the nodes are irreversibly changed. However, there are systems in
nature or society in which nodes can restore a previous state, such
as, a neuron that goes from inactive to active due to external
stimulation. Besides, another simplification of the original model is
that the interactions between networks correspond only to dependency
relations, but it is known that interaction among systems, such as in
ecology, could be trophic, mutualistic or
parasitic~\cite{pocock2012robustness,pilosof2017multilayer}. Therefore
a theory is needed to describe these general relations and reversible
dynamics. Recently, Ref.~\cite{danziger2018dynamic} developed a model
exploring the effect on dynamics of different types of interactions
between two or more networks. The versatility of the model allowed the
authors to explore cooperation and competitive interactions. They
found that these interactions generate rich dynamics, including
chaotic regimes, and hysteresis which are absent in interdependent
networks. Another important simplification of the original model is
that the cascade of failure is based on the topological structure,
ignoring the flow or load between systems which has been recently
incorporated in Refs.~\cite{scala2016cascades,
  brummitt2012suppressing}.

In the following section, we review the literature on the
effect of spatial embedding on the cascade of failures.

\section{Spatial Interdependent Networks}
Although many of the real networks, such as infrastructures are
embedded in space, the interdependent network models described above
did not include spatial embedding. In this section, we will discuss
models which are embedded in space. The embedding implies that nodes
actually have physical locations and thus the link length is
constrained by the geometrical distance between linked nodes
\cite{barthelemy2011spatial,mcandrew2015robustness}. Indeed, the
spatial embedding significantly influences many aspects of a single
network such as, e.g., the different percolation thresholds for two
dimensional square lattices ($p_c\cong 0.59$) and random regular (RR)
networks ($p_c=1/3$) with the same degree ($k=4$) as the square
lattice. It is worth noting, that in most of the early research on
spatial interdependent networks, square lattices were used as the
model choice. This is reasonable since all two-dimensionally embedded
networks with short length links, will be in the same universality
class~\cite{daqing2011dimension} and thus the overall properties and
phenomenology will be similar for all embedded networks. Nevertheless,
we also show that relaxing the lattice structure (see top of Fig. 3),
leads to equivalent results.

\begin {figure}[H]
 \vspace{0.3cm}
 \centering
 \includegraphics[scale=0.65]{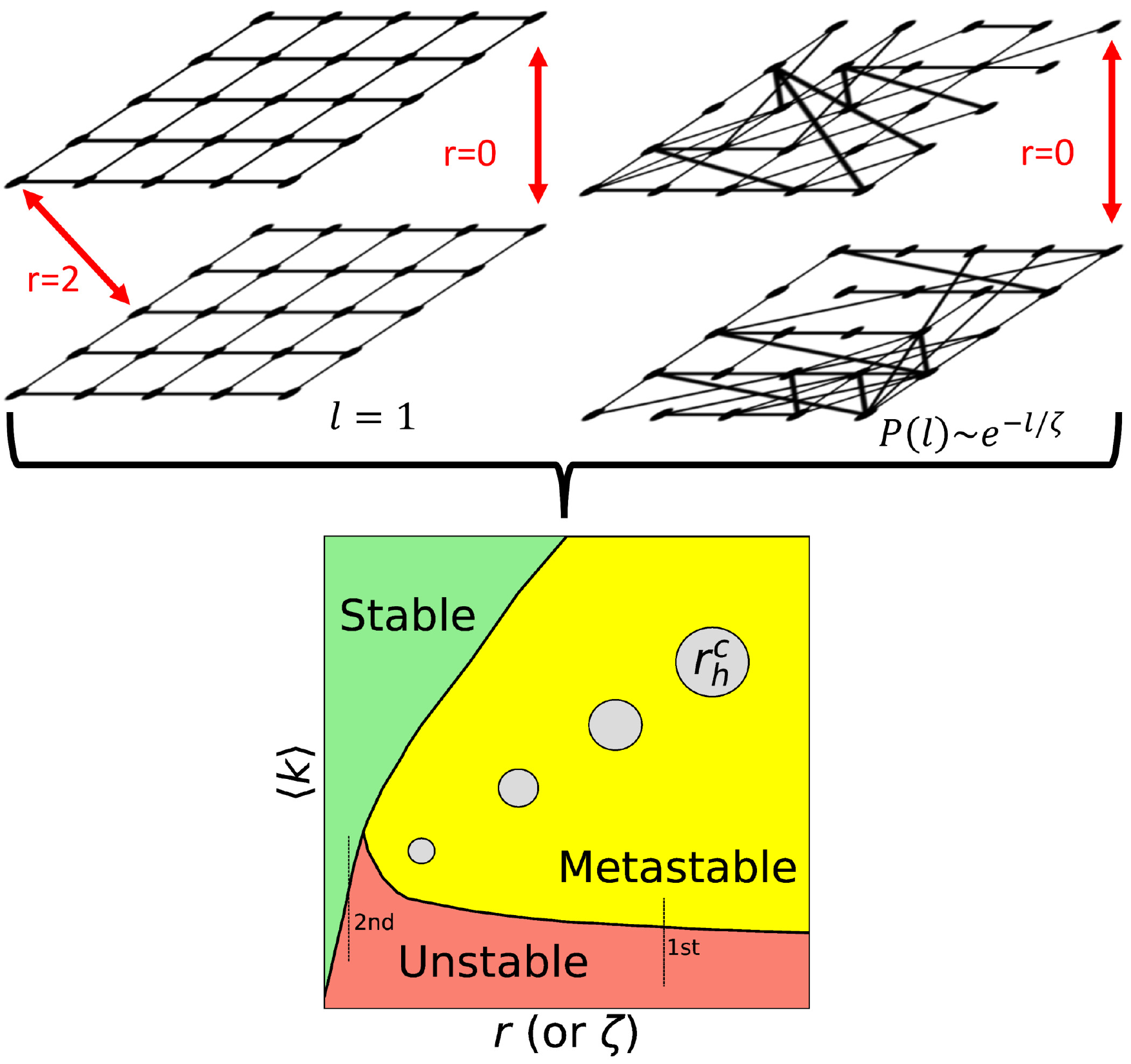}
 \vspace{0.3cm}
 \caption{Upper panel, left: schematic illustration of spatial
   interdependent networks (lattices) with dependency links that connect
   interdependent nodes within a distance $r$. Upper panel, right:
   schematic illustration of spatial interdependent networks with
    connectivity links of length $l$ are taken from an exponential
   distribution, $P(l)\propto \exp(-l/\zeta)$ where $\zeta$
   is the characteristic length of the connectivity links. Lower panel: schematic
   of the obtained phase diagram in the space $\langle k \rangle$ and
   $r$ (or $\zeta$) showing three phases: unstable, stable and
   metastable. In the metastable phase the critical radius $r_h^c$ of
   the localized attack above which the system is destroyed,
   changes. The size of the circles indicates qualitatively that
   $r_h^c$ increases with $\langle k \rangle$ and $r$ (or
   $\zeta$). Furthermore, the two vertical dashed lines represent the
   transitions for random percolation (reducing $\langle k
   \rangle$). For values of $r (\zeta)$ before the peak of the
   unstable red region, the transition is second-order (continuous),
   whereas after the peak of the unstable red region the transition is
   first-order (abrupt). }\label{fig.spatial}
\end{figure}

Incorporating spatial embedding into interdependent networks has been
found to significantly influence the vulnerability of the network. The
first study of such systems was performed in
Ref.~\cite{li2012cascading} analyzing two interdependent square
lattices with a restricted length of dependency links, i.e., a node
$A$ in one network can depend on a node $B$ in the second network
within a distance (between them) smaller than $r$ (see
Fig.~\ref{fig.spatial}). Intuitively, one can realize that if $r=0$
the two networks are identical square lattices (perfectly overlapping)
and thus the percolation properties would be the same as for a single
lattice, i.e., the percolation transition at $p_c=0.59$ is continuous
rather than abrupt. However, Ref.~\cite{li2012cascading} found that
for the case of random initial failures and when $r$ is increased,
there exists a critical length $r_c\approx 8$ above which the
transition becomes abrupt due to the appearance of large
cascades. They also find that at small $r<r_c$ the collapse transition
is continuous, while for large $r>r_c$ it is the abrupt, but the
effect of the maximal vulnerability of the networks at $r=r_c$ and
metastability of the networks for $r>r_c$, when a hole of a sufficient
size destroys the entire network, found in Ref.~\cite{li2012cascading}
is present only below 6 dimensions, suggesting that as in regular
percolation, 6 is the upper critical dimension of this problem. Later
works expand these studies to randomly connected
graphs~\cite{kornbluth2014cascading} in which the length of the
dependency links is defined as a chemical distance, and to high
dimensional lattices~\cite{lowinger2016interdependent}. Similar
model~\cite{son2011percolation,berezin2013comment} with diluted
lattices and dependency links of length $r=0$ also produces a
continuous transition as the model of Ref.~\cite{li2012cascading} does
for $r<r_c$.

Ref.~\cite{bashan2013extreme} considered two interdependent lattices
in the limit $r \to \infty$, i.e., unrestricted length of randomly
chosen dependency links, yet varied the fraction $q$ of interdependent
nodes (with $1-q$ nodes being autonomous). Surprisingly, and in
contrast to non-embedded random networks~\cite{parshani2010interdependent}, it was
found (both analytically and via simulations) that for any level of
interdependence $q>0$ the system undergoes an abrupt, first-order
transition. This is in marked contrast with non-embedded networks
where a critical dependency $q_c>0$ is found.  

 Later work combined the studies of Ref.~\cite{bashan2013extreme} and
 Ref.~\cite{li2012cascading} by varying both $q$ and $r$
 simultaneously~\cite{danziger2014percolation}. There it was found
 that $r_c$ increases as $q$ decreases and that the nature of the
 cascades can be vastly different for $r$ close to $r_c$ as opposed to
 $r\gg r_c$. Namely for $r$ near $r_c$ a cascade front develops and
 spreads radially outward whereas for $r<r_c$ there is no spatial
 propagation front.  Further work, considered $n>2$ interdependent
 networks and found that for $n>11$ even interdependent networks with
 $r=1$ have an abrupt
 transition~\cite{shekhtman2014robustness}.

The above described studies focus on random failures in spatial
interdependent networks, yet the concept of spatiality led to the
recognition that it is possible that an entire network or NONs will
fail in tandem due to a ``localized attack.'' This localized attack or
failure could be due to an earthquake, extreme storm event, or other
reason. The first study on localized
attack~\cite{berezin2015localized} found that, in contrast to random
attack where a finite fraction of the network must fail in order for
the system to collapse, for a localized attack, a finite number of
nodes, constituting a zero fraction of the system (when $N \to
\infty$) can fail, causing cascade and lead to the total system
collapse. It was shown that for certain values of $\langle k\rangle$
and $r$ there exist a metastable state in which a localized attack
with a radius above a critical value $r_c^h$ propagates as a cascade
of failures and yield to collapse of the entire system, while
localized attacks below that size will not propagate. The value of
$r_c^h$ does not depend on the system size but only on the mean
connectivity $\langle k \rangle$ in each network or layer and on the
dependency link length $r$ (see Fig.~\ref{fig.spatial}). In the
$\langle k \rangle-r$ phase space, they also found a stable state in
which there is always a GC for any localized attack size
($r_c^h=\infty$), and an unstable state where the entire system
collapses without any localized attack (see
Fig.~\ref{fig.spatial}). This concept of localized attack was later
extended also for the case of non-spatial networks where it was shown
to often lead to distinct results from random failures~\cite{yuan2015breadth}.

More recent work has exchanged the spatiality roles of the dependency
and connectivity links in spatial interactions. Rather than having the connectivity links
defined by a square lattice with nearest neighbor links of length $1$,
these new models, called $\zeta$-models, have considered connectivity
links with links drawn from an exponential distribution such that
$P(l)\propto \exp(-l/\zeta)$ where $\zeta$ is the characteristic
length of links (see Fig.~\ref{fig.spatial}). Two such sets of links
can be constructed for the same set of nodes, and thus the dependency
links have length zero, representing a spatial multiplex. This is more
realistic since dependency on other infrastructures is more likely to
be on nearby nodes, however connectivity links for supply within a
network can often involve larger distances. In
Ref.~\cite{danziger2016effect}, the authors developed this model and
found that for $\zeta$ below some critical $\zeta_c$ (where $p_c$ is
maximal) the transition is continuous yet for larger $\zeta$ the
transition is abrupt, in strong analogy with the results for $r$ in
Ref.~\cite{li2012cascading} (see bottom of
Fig.~\ref{fig.spatial}). Indeed, a more recent work considered
localized attacks on the $\zeta$-model and found that removing a
finite number of nodes can also lead to total system
collapse~\cite{vaknin2017spreading} (similar to
Ref.~\cite{berezin2015localized}). These two studies have demonstrated
a strong analogy between dependency links of length $r$ with the
connectivity of a spatial lattice and connectivity links from an
exponential distribution with mean length $\zeta$ and dependency links
of length zero.  Both lead to a similar phase diagram, as seen in
Fig.~\ref{fig.spatial}. The reason for this similarity is that each
type of links can bring the failures to distances of $r$ or $\zeta$.

In Ref.~\cite{di2016cascading} the authors used another internal rule
of failure in which nodes survive if they belong to the GC or finite
clusters with sizes greater than $s^*$ because nodes in large finite
components could still receive sufficient resources to continue
functioning. Remarkably, the authors found a region of values of $r$ in which the
failure propagates as moving interfaces that belong to the
universality class of a quenched Kardar-Parisi-Zhang
equation~\cite{kardar1986dynamic}.

\section{Mitigation and Recovery Strategies}

While the problem of cascade of failures has been extensively studied in
the last decade, researchers from different areas have begun recently to
develop strategies to mitigate the effects of the cascades and
avoiding system collapse. These strategies have been focused not only
on mitigation of the cascade but also on the recovery of the
systems. For instance, for overload propagation, some strategies are
based on systematically removing some nodes~\cite{motter2004cascade} or
changing the coupling between systems~\cite{zhang2018cascading}. On
the other hand, for interdependent networks, several strategies
consist of adding connectivity links through a healing
process~\cite{stippinger2014enhancing,stippinger2018universality},
while others increase the functional giant component as the failure
propagates~\cite{di2016recovery}. In this section, we will review some
of the strategies where more details can be found in the cited scientific
literature.

\subsection*{Strategies to Mitigate Overload}

Motivated by the model presented in Ref.~\cite{motter2002cascade},
researchers proposed and studied strategies to mitigate the cascade of
failure due to overloads in single isolated networks~\cite{motter2004cascade}. These strategies consist of intentionally
removing a fraction of nodes just after the initial failure, based on
the smallest load, degree, and closeness centrality. All these
measures are almost equally effective in halting the cascade of
failures because they are highly correlated. The author found that
there exists an optimal value for the fraction of intentionally
removed nodes for which the size of the GC at the end of the cascade
is maximum.

Another proposed strategy~\cite{liu2014modeling} to mitigate the
cascading failure due to overload was based on a spontaneous
self-healing mechanism. In the overload cascade, all the nodes have
the same load tolerance and a node fails when exceed its
capacity. Then, it distributes its load among all of its functional
neighbors. The model starts with the failure of the highest degree
node in the network. Restoration begins at time $t_r$ after
the initial failure and consists in restoring a fraction of the failed
nodes, giving them a new load tolerance higher than the initial
one. The authors found that there exists an optimal restoration timing
at which the size of the functional network is maximum.

In Ref.~\cite{zhang2018cascading} the authors developed a model of two
interacting networks, based on Ref.~\cite{scala2016cascades}, where
the interaction is produced by the load or coupling passing from one
network to the other. In their model, part of the load of a failed
line is sent to the other network through the coupling while the other
part is redistributed within the first network. Even though the
authors do not propose a strategy to avoid the cascade, they studied
an optimal range for the coupling (i.e., the fraction of load in one
network that is distributed to the other networks) of each network
that maximizes the robustness of the system. They found that some
values of coupling between the two networks increase the probability
that both networks survive. An interesting consequence of this model
is that it exhibits multiple discontinuous transitions, but in which
the number of iterations to reach the steady state diverges only in
the final breakdown of the network.

\begin {figure}[H]
\vspace{0.3cm}
\centering
\includegraphics[scale=0.5]{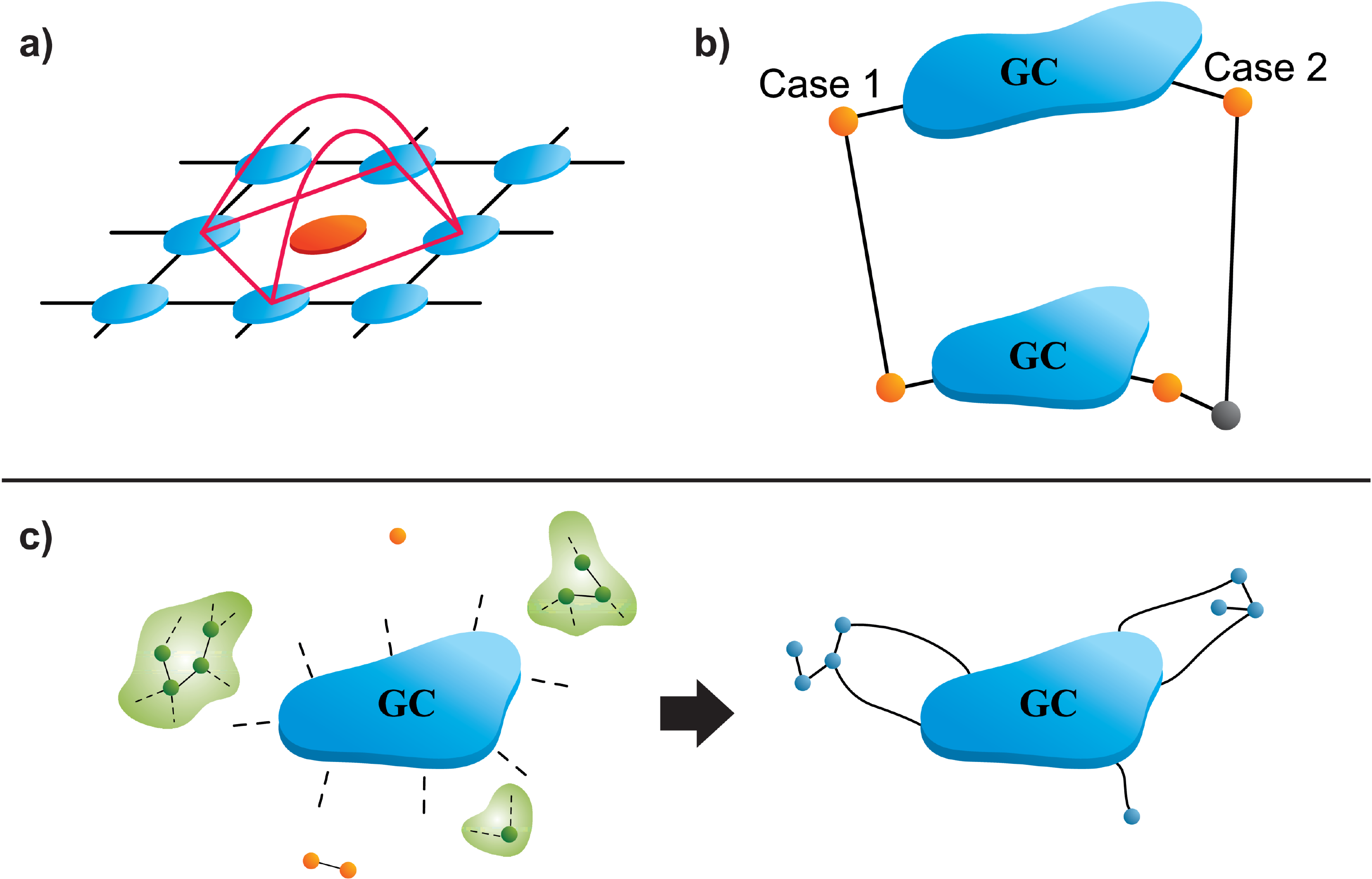}
\vspace{0.3cm}
\caption{Schematic of different strategies for mitigation of
  interdependent networks.  Panel a): healing strategy for
  interdependent lattices proposed in
  Refs.~\cite{stippinger2014enhancing,stippinger2018universality}. The
  healing process consists of creating links (red) with a probability
  $\omega$ between the nearest neighbors of each failed node
  (red). Panel b): schematic of a recovery strategy presented in
  Ref.~\cite{di2016recovery}. The dysfunctional nodes at a distance
  one to their GC are in red, while the rest of the dysfunctional
  nodes at higher distances are in black. If a pair of dysfunctional
  interdependent nodes are at a unit distance from their GC, they are
  restored with probability $\gamma$ (Case 1). However, if one of them
  is not at distance one from the GC, they cannot be restored (Case
  2). Panel c): schematic of a healing strategy for non-embedded
  interdependent networks presented in Ref.~\cite{la2018strategy}. On
  the left, there is a snapshot of the GC and the cluster distribution
  in one layer during the cascade of failure: clusters that have not
  been saved by the strategy are in red, and those that will be
  reconnected to the GC are in green. Dashed links lead to
  dysfunctional nodes (not shown). On the right, some finite clusters
  have been reconnected to the GC.}\label{fig.strat}
\end{figure}

\subsection*{Mitigation by Healing Nodes}

Refs.~\cite{stippinger2014enhancing,stippinger2018universality}
proposed a healing model to mitigate the cascade of failures in
embedded spatial networks. The strategy is applied at each time step
of the cascading failure. The healing process consists of connecting
with a probability $\omega$, the nearest neighbors of each failed node among them
(see Fig.~\ref{fig.strat}a). The idea of this model is to replace the
connections that were lost when the nodes failed. The authors found
that the discontinuous transition can be turned into a continuous
transition depending on the parameter $\omega$. Below a threshold
value $\omega_c$ the nodes connect local nodes and both networks still
preserve the structure of a lattice and the transition is
abrupt. However, for $\omega \geq \omega_c$, the transition is
continuous, the healing promotes the formation of densely connected
regions, connectivity links begin to join distant nodes, and the
structure of the lattice is lost. 

Another strategy that is similar to the healing process is the model
proposed in Ref.~\cite{la2018strategy} which consists in saving finite
clusters before they fail (see Fig.~\ref{fig.strat}c). During the
cascading, each new finite cluster is attached to the GC with a finite
probability. They found that as this probability increases the system
can support a higher level of damage before the functional components
collapse. While the transition is still discontinuous, the jump in the
size of the GC at the transition point is reduced as the probability
to attach clusters increases.

\subsection*{Repair and Recovery Strategies in Interdependent Networks}

Until now, we have discussed strategies to mitigate the effects of the
cascade of failures. However, a few models have been proposed with the
purpose not only to mitigate cascades but also to recover part of the
affected system. In Ref.~\cite{majdandzic2016multiple} the authors
generalized the model of recovery and failure in single networks
developed in Ref.~\cite{majdandzic2014spontaneous} to multiple
networks. They found the minimal fraction of nodes needed to be
repaired in each system to bring the entire system into the functional
state. This set corresponds to the minimal Manhattan distance in the
phase diagram that connects the starting point in which all networks
are dysfunctional to the region where all of networks are
functional. Using a mean field approach, for two interdependent
systems they uncovered a hysteresis and a rich phase diagram with
multiple triple points and four stable solutions following three
scenarios: i) both networks are in high activity, i.e., functionality
ii) both networks are at low activity, and iii) only one network is
functional while the other fails.

Recently, a repairing strategy has been proposed that not only
mitigates the cascade of failures of the model in
Ref.~\cite{buldyrev2010catastrophic} but also can repair a failed
system~\cite{di2016recovery}. The rules for the internal and external
failures are the same as in Ref.~\cite{buldyrev2010catastrophic}. The
strategy consists of repairing with a probability $\gamma$ each pair
of interdependent nodes that belongs to the perimeter of the GC in
each network, where the perimeter is defined as the set of failed
nodes at a chemical distance one from the GC (see
Fig.~\ref{fig.strat}b). The authors found that as $\gamma$ increases,
the system becomes more resilient to the cascade of
failures. Moreover, the authors found that above a critical threshold
of probability of repairing, the system is fully restored.

\section{Conclusions}

Cascading failures is one of the most important processes in complex
networks which show how perturbations induce further failures that
finally could lead to abrupt transitions at a global scale.  The
presence of dependency between different components is crucial for the
propagation of a cascade.  The models of interdependent complex
networks discussed here are more vulnerable than single networks
because fewer failures lead to system collapse. Moreover, spatial
interdependent networks under localized attack can be even more
vulnerable because of the space constraints. In recent years, studies
in this field have increased dramatically, and new avenues of research
continue to be developed, leading to a broader and deeper
understanding of the origin and the effects of cascades of
failures. For example, interacting networks may not only
negatively affect each
other~\cite{leicht2009percolation,brummitt2012suppressing}.  Often,
additional networks are specially built to increase the reliability of
the original networks. For example, SCADA is built to increase the
reliability of the power grid. If the malfunction of SCADA elements
never leads to malfunction of the power grid elements, its presence
can only increase the robustness of the power grid. What makes
interdependent networks more vulnerable is the particular property of
interdependence: the nodes in one network cannot function without
support from the nodes on which they depend. In fact, addition of
various type of protection, such as local control
stations~\cite{rezai2017key}, uninterrupted power supply for SCADA
units, smart grid systems, increases the robustness of interdependent
networks. For example, as pointed out in
Ref.~\cite{korkali2017reducing}, the models of smart grid in which the
failure of the control units of the communication network does not
lead to the failure of a power station which they serve, the
robustness increases with coupling strength between the power grid and
communication network.  However, if the failure of the communication
node leads to the failure of the power node, the robustness decreases
with the coupling strength in agreement with the predictions of
topological models, e.g.~\cite{parshani2010interdependent}. In any
case, analysis of real catastrophic blackouts demonstrates that
failures of the control cyber-systems often plays the crucial role in
their development~\cite{rosato2008modelling,nyiso2004}. Further
examples of very recent studies include understanding the stability of
the GC~\cite{kitsak2018stability}, finite sizes
effects~\cite{zhou2014simultaneous,radicchi2015percolation}, and
developing strategies to avoid an abrupt collapse of the
system~\cite{la2018strategy,stippinger2018universality}.

Although several lines of research have provided an underlying
framework, there are still many open questions that remain to be
addressed. For instance, one of the significant problems is the lack of real
data on multilayer networks that could be used to develop more
realistic models for cascades of failures. Therefore, the construction
of inference methods such as those suggested in
Ref.~\cite{lacasa2018multiplex}, are necessary to advance our
knowledge. Another problem is that most of the works in
interdependent networks rely heavily on the network structure, but
functionality can be based on the dynamics, such as in
Refs.~\cite{danziger2015interdependent,danziger2018dynamic}.  This
dynamic interdependency research is still in its infancy, and
deeper studies should emerge. In this direction, it
would be particularly significant, for example, to include
interdependencies in the models proposed in
Refs.~\cite{schafer2018dynamically,yang2017cascading}, which would
provide a more comprehensive elucidation of these
processes. Finally, regarding strategies to contain and mitigate a
cascade of failures, it would also be relevant to explore the optimal
time and location to apply such strategies since resources are typically limited.

\section*{Funding}

This work was supported by the Defense Threat Reduction Agency [Grant
  no. HDTRA1-14-1-0017]; UNMdP and CONICET [PIP 00443/2014]; and the
National Science Foundation [Grant no. PHY-1505000].

\section*{Acknowledgments}
The authors thank
Prof. H. Eugene Stanley and Prof. H\^enio Henrique Arag\~ao R\^ego for
useful discussions.

\fbox{\begin{minipage}{40em} 
    {\bf Non-embedded and embedded networks in an Euclidean space}

There exist real systems in which the topology and the process that
develop on top of them are not constrained by the Euclidean distance,
such as in the World Wide Web or Facebook. These systems are better
modeled by non-embedded networks, and the simplest case is the
Erd\H{o}s-R\'enyi graph. One of the most remarkable properties of these
systems is that they are small world, i.e., the chemical distance or
hop count between two nodes increases very slowly with the system
size, as $\ln{N}$ or $\ln{\ln{N}}$. On the other hand, there are
networks in which the topology is constrained by the space since nodes
cannot connect to others that are spatially too far away, such as the road
network between junctions or cities. The simplest model of an embedded network is a
lattice. In these systems, the average distance increases faster than
in non-embedded networks, as $N^{1/d}$ where $d$ is the dimension of
space of the embedded network. The small-world property in
non-embedded networks implies that different processes on top of these
networks such as a cascade of failures or the propagation of a rumor
are much faster than in a Euclidean lattice.
\end{minipage}}

\fbox{\begin{minipage}{40em} 
    {\bf Glossary}
    \begin{itemize}
      \itemsep0em
    \item Cascade of failures: Propagation of a failure through the
      system where a component that becomes dysfunctional lead to other
      components that depend on it (directly or indirectly) to also become
      dysfunctional.
    \item Giant Component (GC): A cluster with a macroscopic number of nodes.
    \item Finite cluster: A cluster with a microscopic number of nodes.
    \item Connectivity link: A link that connects two nodes from the same network.
    \item Dependency link: A link that connects a pair of nodes such
      that if one fails, then the other one also fails, even if still
      connected to the GC.
    \item Autonomous node: A node in an interdependent network that
      does not need any dependency link.
    \item Continuous phase transition: A phase transition such that at
      the critical threshold the order parameter (for example, the
      magnetization in the Ising model, or the size of the GC in
      percolation) approaches to zero continuously and many properties
      behave as power laws when approaching this threshold. If
      the derivative of the order parameter is discontinuous, then the
      transition is of second order.
    \item Discontinuous phase transition: A phase transition in which
      at a threshold value the order parameter is discontinuous and
      changes abruptly. It is usually called a ``first order
      transition.''
    \item Random failure, targeted and localized attack: Nodes can
      fail or be removed initially using different rules: at random
      (random failure), according to their degree or their centrality
      (targeted attack), or in nearby shells around a node (localized
      attack).
    \item Clustering: It is a measure of the density of triangles.
    \item Degree-degree correlation: Related to the probability that a
      node with degree $k$ is connected to a node with a similar or
      different degree.
      \end{itemize}      
\end{minipage}}
\fbox{\begin{minipage}{40em} 
    {\bf Glossary}
    \begin{itemize}
    \item Modular network: A network composed of groups with a high
      internal density of connectivity links and a sparse density of
      connectivity links between these groups nearby (called modules
      or communities). The nodes that connect between two or more
      communities are called ``bridge nodes.'' Modular networks can be
      found, for example, in metabolic systems, neural networks,
      social networks, or
      infrastructures~\cite{ravasz2002hierarchical,happel1994design,gonzalez2007community,eriksen2003modularity}.
    \item Betweenness centrality: The fraction of shortest paths
      between all pair of nodes passing through a given node or a
      link~\cite{mej2010networks}.      
    \item $k$-shell: It is obtained by removing, iteratively, all nodes
      with degree smaller than $k$, until all remaining nodes have
      degree $k$ or larger~\cite{carmi2007model}.
    \item Hierarchical network: a modular network in which each module
      is formed by other modules, and so on. For example, countries could
      represent modules and, at the same time, each province, state or
      region in each country could also have modules or sub-modules.
    \end{itemize}
\end{minipage}}

\fbox{\begin{minipage}{40em}
{\bf Percolation in single networks }

Percolation theory studies the statistical properties of clusters of
nodes generated when links or nodes are removed from a network. Many
physicists and mathematicians made contributions to this field; showing how percolation
can describe different phenomena on Euclidean networks such as forest
fires, the spread of diseases, and electric conduction in disordered
environments~\cite{araujo2014recent}. Currently, the theory of
percolation has had a boom with the emergence of complex networks,
since it allows researchers to describe and understand phenomena and
processes such as the robustness of very heterogeneous networks
against random failures and the difficulty of eliminating a virus from
the Internet
network~\cite{cohen2010complex,araujo2014recent,boccaletti2006complex}. One
of the most studied percolation processes is random node or link
percolation, in which a fraction $1-p$ of nodes or links are removed
randomly. This process exhibits a second order transition at a
critical value $p=p_c$ above which a GC exists, while below $p_c$ the
system contains only finite clusters. Other percolation processes that
lead to cascades are k-core and bootstrap
percolation~\cite{dorogovtsev2006k,baxter2010bootstrap,baxter2011heterogeneous,di2018insights,carmi2007model}. The
former is used to study the deactivation process in which a fraction
of nodes fail initially and then a node fails or becomes dysfunctional
if it has a number of active neighbors below a chosen threshold. On the other
hand, bootstrap percolation is an activation process, in which a
fraction of nodes are initially activated, and then a node becomes
functional if the number of active neighbors is above a threshold.
\end{minipage}}

\bibliographystyle{comnet}
\bibliography{mibibDoc}

\end{document}